\documentclass[pre,twocolumn,showpacs]{revtex4}
\usepackage{graphicx}

\newcommand{\equ}{Eq.}

\newcommand{\fig}{Fig.}

\newcommand{\rem}[1]{}

\begin{document}

\title{Evanescent wave approach to diffractive phenomena in convex billiards
with corners} 
\author{Jan Wiersig, Gabriel G. Carlo}
\affiliation{Max-Planck-Institut f\"ur Physik komplexer Systeme, D-01187
Dresden, Germany}
\date{\today}
\email{jwiersig@mpipks-dresden.mpg.de, carlo@mpipks-dresden.mpg.de}

\begin{abstract} 
What we are going to call in this paper ``diffractive phenomena" 
in billiards is far from being deeply understood. These are  
sorts of singularities that, for example, some kind of corners introduce 
in the energy eigenfunctions.
In this paper we use the well-known scaling quantization procedure 
to study them.
We show how the scaling method can be applied to convex billiards 
with corners, taking into account the strong diffraction at them and 
the techniques needed to solve their Helmholtz equation. 
As an example we study a classically pseudointegrable billiard, the truncated
triangle. Then we focus our attention on the spectral behavior.  
A numerical study of the statistical properties of high-lying energy levels
is carried out. It is found that all computed statistical quantities are
roughly described by the so-called semi-Poisson 
statistics, but it is not clear whether the semi-Poisson statistics is the
correct one in the semiclassical limit. 
\end{abstract}
\maketitle

\section{Introduction}

In this work we calculate very high lying eigenvalues of 
a billiard system using the so-called scaling method~\cite{VerginiSaraceno95, 
PhD1,PhD2}.
This method has two main advantages, it is formulated on the boundary of 
the billiard (allowing matrices of order $k$, the wavenumber) and 
it avoids zeroes searching algorithms.
We have solved the Helmholtz equation with Dirichlet boundary 
conditions on the billiard boundary. A point worth to 
mention is that any eigenfunction is $C^{\infty}$ at domain points. 
At straight segments of the boundary, eigenfunctions are reflected as 
odd functions, so that the result satisfies the Dirichlet condition. 
Though being also $C^{\infty}$ at the boundary by straight pieces, 
they are not analytical at a vertex when two segments of the boundary join 
(with inner angle $\pi/r$, where $r$ is a noninteger number). 
This feature leads to what we address as diffractive phenomena. 

The main idea behind the scaling method is that the trial functions 
of the variational problem can be parametrised by energy (alternatively, 
by the wavenumber $k$). With this idea in mind these 
functions are expanded in a suitable scaling basis (plane 
waves with different propagation directions are one example and 
these are the elements we use here). Asking the function to be zero 
at the boundary of the billiard is the same as asking its norm there 
to be zero as well.  
Then, eigenvalues (and eigenfunctions) can be obtained by solving a 
generalised eigenvalue problem that involves the quadratic form 
associated with the norm of the function on the boundary. 

In applying the method to this system, evanescent waves were 
needed. This is because real plane wave solutions to the Helmholtz 
equation cannot represent all the features of diffraction. 
It has been shown~\cite{Berry94} that 
an evanescent plane wave, which oscillates along propagation direction faster 
than the wavenumber $k$, can be constructed by means of real plane waves. 
Nevertheless, the corresponding superposition is a singular one, suggesting 
the direct use of the evanescent 
functions in the basis. This is the way we have dealt with diffraction.  

\begin{figure}[ht]
\includegraphics[width=5.0cm,angle=0]{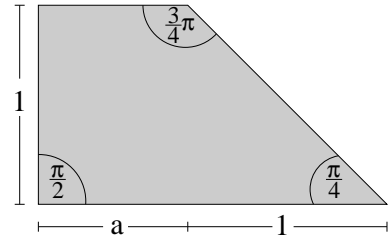}
\caption[]{\footnotesize The truncated triangle.}
\label{fig:billiard}
\end{figure}

For simplicity we restrict our considerations to an angle of $3\pi/4$. A
particularly suitable example to study the effect of this kind of corner on the
quantum properties is the polygonal billiard shown in
\fig~\ref{fig:billiard}. This billiard is called the truncated triangle. It
has been studied in a variety of  
contexts~\cite{RichensBerry81,HP82,SSSSSZ94,SimmelEckert95,SimmelEckert96,
KudrolliSridhar97,AGR00}. 

The truncated triangle belongs to the class of {\it rational polygons}. That
are polygons where all angles $\alpha_j = m_j\pi/n_j$
between sides are rationally related to $\pi$, where $m_j, n_j > 0$ are
relatively prime integers.   
The free motion inside a rational polygon is integrable if $m_j = 1$ for all
$j$, which is the case for rectangles, the equilateral triangles, the $\pi/2,
\pi/4,\pi/4$-triangles and the $\pi/2, \pi/3, \pi/6$-triangles. All other
rational polygons are {\it pseudointegrable}~\cite{RichensBerry81}. Like in
integrable systems, the phase space is foliated by two-dimensional invariant
surfaces~\cite{Hobson75,ZemlyakovKatok75}. However, the genus of the surfaces
is greater than one due to {\it critical corners} with $m_j >
1$~\cite{RichensBerry81}. In the case of the truncated triangle, the genus is
$g=2$. Roughly speaking, the invariant surface is a torus with an additional
handle. 

Quantum signatures of pseudointegrability can be found in the energy 
eigenfunctions~\cite{Wiersig01} and in the statistical properties of energy
levels~\cite{RichensBerry81}. The energy levels of pseudointegrable systems are
correlated in contrast to those of integrable systems which are
generically well described by the Poissonian random 
processes~\cite{BerryTabor77a}.  
For example, the nearest-neighbor spacing distribution of pseudointegrable
systems generically displays level repulsion~\cite{RichensBerry81},
resembling the Gaussian orthogonal ensemble (GOE) of random-matrix
theory~\cite{Mehta67} which describes fully chaotic systems with time-reversal
symmetry~\cite{BGS84}. Significant deviations from GOE have been
observed first in Refs.~\cite{SS93,SSSSSZ94}. 
It has been suggested that the spectral statistics of pseudointegrable
systems is another example of critical or intermediate
statistics~\cite{BGS99,BGS01b}. Critical statistics appear in many condensed
matter problems such as in
mesoscopic disordered systems at the critical point of the metal-insulator 
transition~\cite{SSSLS93}, in systems with a few interacting
electrons~\cite{WWP99}, and in incommensurate multiwalled carbon
nanotubes~\cite{AKWC02}.  

In Refs.~\cite{BGS99,BGS01b} it has been proposed to use the semi-Poisson (SP)
statistics as a reference point for critical statistics. The SP statistics is
defined by a simple construction: remove every other level from an ordered
Poisson sequence~\cite{HFS99,BGS01b}. The SP statistics is useful because it
provides explicit formulas for a number of statistical quantities which can be
compared to the statistical properties of a given system.
For several pseudointegrable
systems~\cite{BGS99,BGS01b,GremaudJain98,Gorin01,PG01,BGS01,Wiersig02} it has
been confirmed that the SP statistics indeed describes the short-range level
correlations rather well. Comparing the long-range level correlations is
numerically a difficult task because the statistical properties of rational
polygons converge extremely slowly as energy is
increased~\cite{BGS01,Wiersig02}. 
Fortunately, semiclassical periodic-orbit theory allows to compute
analytically the long-range level correlations, in terms of the level
compressibility $\chi$, of a few special systems, like certain right
triangles~\cite{BGS01} and the barrier billiard~\cite{Wiersig02}. In the
former case $\chi$ differs in general from the SP result, whereas in the
latter case $\chi$ is in agreement with the SP statistics. Nothing is known
analytically about the generic case.

Our numerical analysis will show that the truncated triangle is well
described by the SP statistics but deviations are not negligible. Our analysis 
extends that reported in Ref.~\cite{SSSSSZ94} in many
respects: (i) The scaling method allows us to compute more levels,
giving a better statistics. (ii) Moreover, high-lying energy levels can be
computed. This puts us in a position to study the relevant asymptotic regime. 
(iii) More statistical quantities are computed.
(iv) The numerical results are compared to the SP statistics.

The paper is organized as follows: in Section \ref{sec:scaling}, we describe 
the features that are considered when applying the scaling method 
to the system under investigation. In Section \ref{sec:numerics}, we present 
the statistical studies carried out with the data that has been 
obtained. Finally, Section \ref{sec:conclusion} is devoted to conclusions. 

\section{The scaling method applied to the truncated triangle}
\label{sec:scaling}
We are not going to explain the scaling method here, and we address 
the reader to the Appendix and the given references for details. 

An eigenfunction of a billiard can be constructed as a plane wave 
superposition. This can include evanescent waves, i.e., plane waves 
with complex wave vectors. These types of waves should be present in quantum 
billiards and can be associated with diffractive phenomena.  
Discontinuities at the boundaries seem to be strongly related to 
the way evanescent functions must be considered to solve the problem. 

Several authors have focused their attention on this 
issue~\cite{PhD1,APS91,Heller91,VUF95,Berry94}. 
For polygonal billiards, in the generic case, theory suggests that 
there is no real plane wave superposition that can be an 
eigenfunction~\cite{APS91}. On the other hand, good numerical results using 
only plane waves could be found~\cite{VUF95}, but working 
within the region of low energies. Evanescent waves were studied 
in detail by Berry~\cite{Berry94} in the context of quantum billiards. 
In this approach, the main reason to do it comes from the 
idea of constructing them by continuation of an external 
scattering superposition, containing only real plane waves. 
In principle, this does not seem to be possible. Anyway, he showed 
how evanescent waves can be expressed as the singular limit 
of an angular superposition of real plane waves. 

In our system, 
and working at high energies, we need to consider evanescent 
waves explicitely, because a real plane 
wave representation is singular. In the semiclassical 
limit, the only way to obtain eigenfunctions that include 
evanescent waves is by considering them in the variational 
problem~\cite{PhD1}. In the remaining part of this section we are 
going to show some examples of the waves considered and the 
idea behind their selection. 

Plane wave solutions to the Helmholtz equation can be written 
as 
\[
\psi({\bf r}) \; = \; \exp{[{\it i}k \cos{(\theta + {\it i} 
\alpha}) x + {\it i} k\sin{(\theta + {\it i} 
\alpha}) y]}
\]
with $\theta$ and $\alpha$ real, using coordinates ${\bf r} 
= (x,y)$ in the plane. We can express this wave in a 
slightly different fashion by distinguishing the real and 
imaginary parts in the exponent, 

\[
\psi({\bf r}) \; = \; \exp{[{\it i}k \cosh{(\alpha)}\tilde x ]} 
\; \exp{[-k \sinh{(\alpha})\tilde y]}. 
\]
Here, the propagation direction $\tilde x= x \cos{\theta} + 
y \sin{\theta}$ implies an angle $\theta$ over the $x$ axis. 
In this direction the 
function has a wavelength that is given by 
$2 \pi / (k \cosh{\alpha}) < 2 \pi / k$. However, in the 
orthogonal direction $\tilde y$ the function is an 
exponential with 
coefficient $-k \sinh{\alpha}$. 

\begin{figure}[ht]
\includegraphics[width=6.0cm,angle=0]{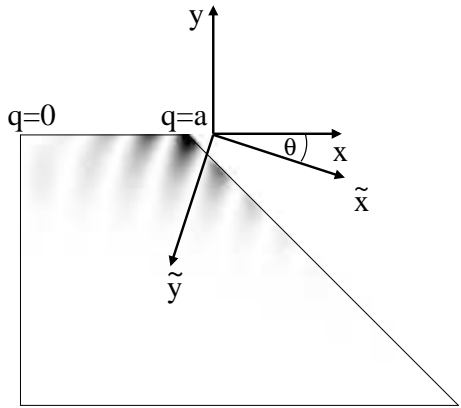}
\caption[]{\footnotesize Evanescent wave for $k=20$ on 
the truncated triangle domain with
parameter value $a=(\sqrt{5}-1)/2$. The propagation 
direction $\tilde x$ (given by $\sin{\theta} = -0.3$ 
in this case) and the corresponding decaying direction 
$\tilde y$ are both shown by the arrows. Also, we point 
out values $0$ and $a$ of the arclength coordinate $q$.}
\label{evadom}
\end{figure}

An example of these functions applied to our system 
can be found in \fig~\ref{evadom}, where we show an 
evanescent wave corresponding to the truncated triangle 
of parameter value $a=(\sqrt{5}-1)/2$ (Note that since the absolute size of
the system is irrelevant, we measure length scales as dimensionless 
quantities). In this case we 
chose $k = 20$ in order to have a visible wave. For the 
higher-lying wave numbers ($k \simeq 1000$ or greater) 
these functions are difficult to see in the domain since 
they decay very fast. 
Nevertheless, the slower oscillating functions 
we have used in the latter energy region 
can be seen over the boundary.

\begin{figure}[ht]
\includegraphics[width=7.0cm,angle=0]{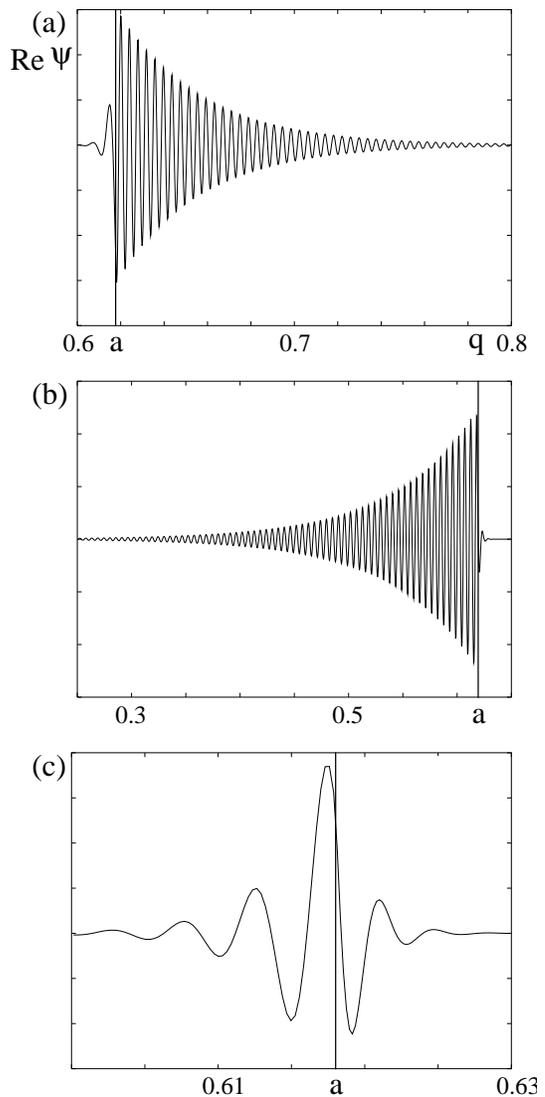}
\caption[]{\footnotesize Three examples of evanescent 
waves over the billiard boundary for $k = 1000$. 
Vertical axis corresponds 
to the real part of the evanescent wave in arbitrary units. 
Horizontal axis 
corresponds to the boundary arclength coordinate $q$ (dimensionless).
In (a) we show a wave decaying 
to the right (propagation direction given 
by $\sin{\theta} = -0.682$), in (b) one to the left 
($\sin{\theta} = -0.025$), 
and in (c) to both sides ($\sin{\theta} = -0.3$) from 
$q=a$, the position of the critical corner.}
\label{evbound}
\end{figure}

Now that the general ideas related to evanescent waves 
have been exposed, we are going to explain the way we 
selected them, and we are also going to show some examples 
for $k = 1000$. The first thing to point out is that 
diffraction is generated at the critical corner with angle $3 \pi / 4$  
that is shown in \fig~\ref{fig:billiard}. So, our sets or 
families of evanescent waves are centered at this corner.
We consider three different kinds of functions, all 
sharing the previous property, but decaying to 
one or the other ``side" of the corner (in terms 
of the arclength boundary coordinate $q$) in the first 
place and also an additional one that decays to both sides. 
We use a family of 12 waves whose propagation direction $\theta$ 
is slightly smaller than $- \pi / 4$, i.e., around the right hand 
side of the boundary as seen from \fig~\ref{fig:billiard}, 
another set of 8 waves with almost horizontal propagation and 
finally one that goes along an intermediate direction 
(for this one we took a value of $\sin \theta= -0.3$, 
as can be seen in \fig~\ref{evadom}). 
In fact, in order to appreciate their contribution, 
it is convenient to look 
at them on the boundary of the billiard. For this 
reason we show in \fig~\ref{evbound} three examples of evanescent 
waves that are good representatives of these three 
families. 
As already mentioned, the exponential decay is from 
the critical corner, which corresponds to $q=a$ in 
terms of the boundary coordinate. This is 
the effect of considering an angle slightly different from 
$- \pi /4$ angle for the propagation direction. The 
same happens with the second family whose example is directed 
almost along the horizontal (left) segment of the boundary. 
Finally, we show the evanescent wave that decays to 
both sides of the corner. 

This approach to the problem has proven to be very efficient. 
As a matter of fact, studying carefully several billiard 
eigenfunctions that show the greatest norm (error) over 
the boundary (without considering evanescent waves in 
their calculation), we could check that these are the 
main components of the diffracted field. 

Then we apply the symmetries of our system in order 
to get the right contributions. This is easy to implement 
by using the symmetry operations of the $C_{2v}$ group. 

We have taken only up to 21 evanescent waves in order 
to get our results. This is a small number compared 
to the roughly 1000 real plane waves that are 
used in the highest-lying energy window we have 
obtained~\cite{VerginiSaraceno95}. But they are key to lower the error 
of the eigenvalues. This shows that, even though the 
relevance of this contribution goes to zero in the 
semiclassical limit we need to consider it in order 
to resolve individual states.

\section{Spectral statistics}
\label{sec:numerics}
We here examine the spectral statistics of the quantized truncated
triangle. We consider the generic case 
where $a$ is an irrational number. In the nongeneric case of $a$ being 
rational, the energy spectrum contains a subset of Poisson distributed
levels~\cite{RichensBerry81}. We focus on the parameter value
$a=(\sqrt{5}-1)/2$, the reciprocal of the golden mean.  For other parameter
values of $a$ [$(\sqrt{5}-1)/2+0.2$, $2/\pi$, and $2/\pi+0.1$] similar
results have been obtained.
We use energy windows of length $20\,000$ in five different regimes
starting with level number $853$, $32\,124$, $89\,607$, $149\,879$, and
$190\,356$ corresponding to the wave number $k=100$, $600$, $1000$, $1300$,
and $1500$. 

To study the local fluctuations in the level sequence  $E_1 \leq E_2
\leq E_3\leq\ldots$ it is necessary to remove the systematic global
energy dependence of the average density. To do so, we ``unfold'' the spectra
in the usual way by setting $\tilde{E}_n = 
\bar{N}(E_n)$; see, e.g., Ref.~\cite{Haake91}. $\bar{N}(E)$ is the smooth part of
the integrated density of states, i.e., the number of levels up to
energy $E$.   
We approximate $\bar{N}(E)$ by the generalized Weyl's law including perimeter
and corner corrections~\cite{BH76}
\begin{equation}
\bar{N}(E) = \frac{A}{4\pi}E-\frac{L}{4\pi}\sqrt{E}+C \ ,
\end{equation}   
where $A=a+1/2$ is the area of the billiard, $L=2a+2+\sqrt{2}$ is the
perimeter, and $C=11/36$ is the corner correction.
The unfolded spectra $\{\tilde{E}_n\}$ have unit mean level
spacing and are dimensionless. Henceforth, the tilde will be suppressed.   

\subsection{Nearest-neighbor spacing distributions}
The most popular statistical quantity in the field of quantum chaos is the 
nearest-neighbor spacing distribution. It is defined as the probability
density of the spacing $s$ (in units of the mean level spacing) between adjacent levels,
\begin{equation}
P(s) = \lim_{n\to\infty}\frac{1}{n}\sum_{i=1}^n\delta(s-E_{i+1}+E_i) \ .
\end{equation}
Clearly, the nearest-neighbor spacing distribution is a measure of short-range
level correlations. We will compute its integral, the cumulative spacing
distribution 
\begin{equation}\label{eq:cnn}
I(s) = \int_0^s P(s')ds' \ .
\end{equation}
For the Poisson statistics $P_{\text{\footnotesize P}}(s) = \exp{(-s)}$ and 
$I_{\text{\footnotesize P}}(s) = 1-\exp{(-s)}$, the GOE is well described by 
the Wigner surmise $P_{\text{\footnotesize W}}(s) = (\pi/2)s\exp{(-\pi s^2/4)}$
and $I_{\text{\footnotesize W}}(s) = 1-\exp{(-\pi s^2/4)}$, and for the SP 
statistics~\cite{HFS99,BGS01b}   
\begin{equation}\label{eq:spnn}
P_{\text{\footnotesize SP}}(s) = 4s e^{-2s}, \,\, 
I_{\text{\footnotesize SP}}(s) = 1-(2s+1)e^{-2s} \ .
\end{equation}
At small $s$, $P_{\text{\footnotesize SP}}(s)$ exhibits a linear increase from
zero (level repulsion) similar to the 
Wigner surmise.  At large $s$, $P_{\text{\footnotesize SP}}(s)$ has an
exponential fall-off as the Poisson statistics.  

\begin{figure}[ht]
\includegraphics[width=7.5cm,angle=0]{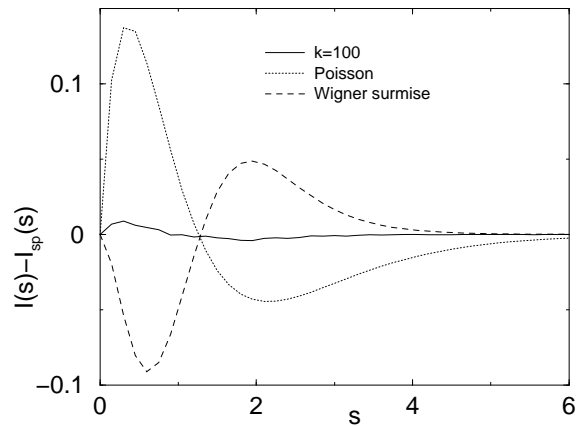}
\caption[]{\footnotesize Difference between the cumulative spacing
distribution of the first energy window and the SP result.}  
\label{fig:cnnzero}
\end{figure}

\begin{figure}[ht]
\includegraphics[width=7.5cm,angle=0]{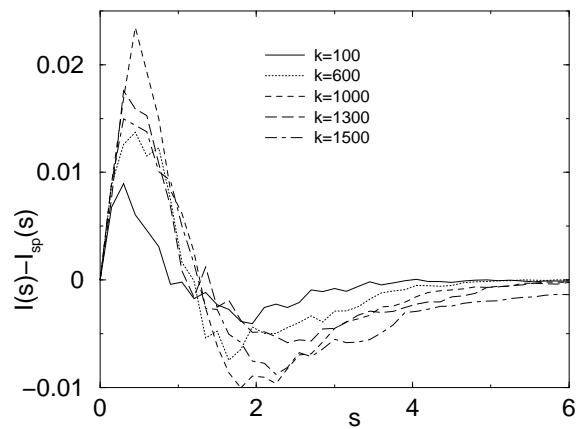}
\caption[]{\footnotesize The cumulative spacing distribution of all energy
windows.}  
\label{fig:cnn}
\end{figure}

In \fig~\ref{fig:cnnzero} we plot the difference between the cumulative spacing
distribution to the SP result for the lowest-energy window. Good agreement
with the SP statistics can be observed. Figure~\ref{fig:cnn} shows a
magnification, containing also the other energy windows. We see increasing
deviations from SP for $k=600$ and $k=1000$, and then decreasing deviations
for $k=1300$ and $k=1500$. No clear trend to SP is visible as one goes to
higher energies.   

Similar small deviations from the SP statistics have been observed at the
metal-insulator transition in the three-dimensional Anderson model. In that
case, an ensemble average over certain boundary conditions removes the
deviations considerably~\cite{BMP98}. To see whether this is possible also in
our case, we consider Neumann boundary conditions on the
two boundary segments which do not touch the critical corner; see
\fig~\ref{fig:billiard}. In this way we obtain four energy spectra
corresponding to Dirichlet/Dirichlet, Dirichlet/Neumann, Neumann/Dirichlet,
and Neumann/Neumann-boundary conditions. In contrast to the case of the
Anderson model we find that averaging over these boundary conditions does
not reduce the deviation from the SP statistics.

\subsection{Next-to-nearest spacing distributions}
We now consider the next-to-nearest spacing distribution 
(second-neighbor-spacing distribution) and its integral. 
The SP statistics gives~\cite{BGS01b}   
\begin{eqnarray} 
P_{\text{\footnotesize SP}}(2,s) & = & \frac83s^3 e^{-2s}\,, \nonumber\\
I_{\text{\footnotesize SP}}(2,s) & = & 1-\frac13(4s^3+6s^2+6s+3)e^{-2s} \ .
\end{eqnarray}

\begin{figure}[ht]
\includegraphics[width=7.5cm,angle=0]{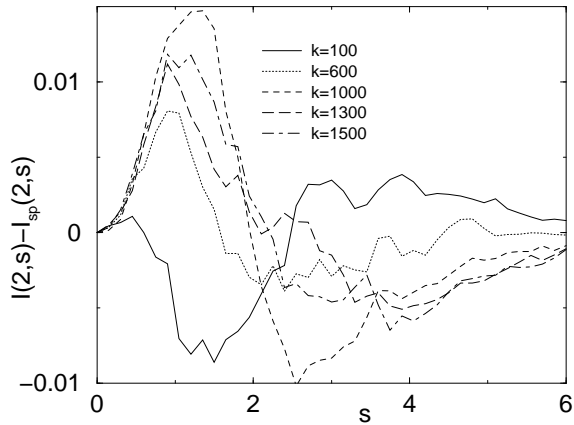}
\caption[]{\footnotesize The cumulative next-to-nearest
spacing distribution.} 
\label{fig:nextcnn}
\end{figure}

Figure~\ref{fig:nextcnn} shows that the cumulative next-to-nearest spacing
distribution is well described by the SP statistics. The maximal deviations
from SP are smaller than in the case of the nearest-neighbor spacing
distribution in \fig~\ref{fig:cnn}. But, again, there is no clear
convergence to the SP statistics as one goes to higher energies.   

\subsection{Number variance}
So far we have studied short-range level correlations (nearest and
next-to-nearest spacing distributions). Long-range level
correlations are conveniently studied with the help of the number variance
\begin{equation}\label{eq:sigma}
\Sigma(L) = \left<(n(L,E)-L)^2\right> \ .
\end{equation} 
$\Sigma(L)$ is the local variance of the number $n(L,E) = N(E+L/2)-N(E-L/2)$ of
energy levels in the interval $[E-L/2,E+L/2]$. For the SP statistics
we have~\cite{BGS99,HFS99,BGS01b} 
\begin{equation}
\Sigma_{\text{\footnotesize SP}}(L) = 
\frac{L}{2}+\frac{1}{8}(1-e^{-4L}) \ .
\end{equation}

\begin{figure}[ht]
\includegraphics[width=7.5cm,angle=0]{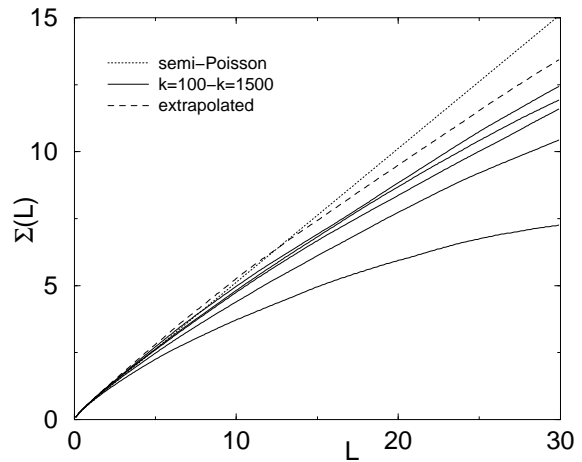}
\caption[]{\footnotesize Number variance $\Sigma(L)$ for $k=100$,
$600$, $1000$, $1300$, and $1500$ (from below).}
\label{fig:sigma}
\end{figure}

Figure~\ref{fig:sigma} shows the number variance computed for the five energy 
windows. Note that the regime is well below the crossover region where
$\Sigma(L)$ begins to saturate at a value determined by the shortest
periodic orbit~\cite{Berry85b}. With increasing energy the number variance
comes closer to the SP result, without showing a clear stabilization. To
estimate the limit curve as $k\to\infty$ we use the extrapolation procedure
described in Ref.~\cite{BGS01}: extrapolate point by point (with $L$ fixed) the
four highest curves with a fit $A(L)+B(L)/k$. The limit curve, $A(L)$, is
shown as dashed curve in \fig~\ref{fig:sigma}. It is closer to the SP
result. To measure the difference we concentrate on the slope in the region 
of large $L$, the so-called level compressibility
\begin{equation}\label{eq:chi}
\chi = \lim_{L\to\infty} \frac{\Sigma(L)}{L} \ .
\end{equation}
We get $\chi \approx 0.41$ from the extrapolated curve. This is close to the
SP result~\cite{BGS01} of $1/2$ ($1$ for Poisson~\cite{Berry85b} and
$0$ for GOE~\cite{Berry85b,Mehta67}).

\subsection{The form factor}
Another measure of long-range level correlations is the form factor $K(\tau)$,
the Fourier transform of the two-point
correlation function. The limit $\tau\to 0$ is related to the number variance
by means of $K(0) = \chi$ (see, e.g., Ref.~\cite{BGS01}) with $\chi$ from
\equ~(\ref{eq:chi}).

\begin{figure}[ht]
\includegraphics[width=7.5cm,angle=0]{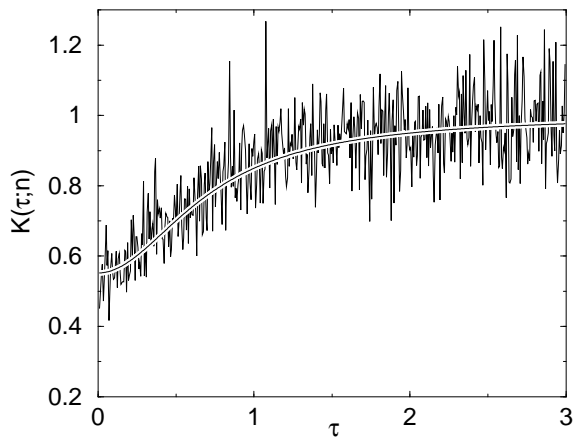}
\caption[]{\footnotesize The form factor~(\ref{eq:Ktn}) for the highest-energy
window averaged over small intervals of size $\Delta\tau = 0.006$. The smooth
curve is the fit~(\ref{eq:Kfit}) with $c=4.464$.}
\label{fig:Kav}
\end{figure}

The form factor can be approximated numerically by (see, e.g.,
Ref.~\cite{Marklof98})  
\begin{equation}\label{eq:Ktn}
K(\tau;n) =  \frac{1}{n}\left|\sum_{j=l}^{l+n} e^{2\pi i E_j\tau}\right|^2 .  
\end{equation}
In our case $n=20\, 000$. Note that $\tau$ is dimensionless. Figure~\ref{fig:Kav} shows $K(\tau;n)$ averaged over
small intervals of size $\Delta\tau = 0.006$ in the  high-energy regime,
i.e., $k = 1500$. It is difficult to estimate 
$K(0)$ from such kinds of noisy data, but it is justified to say that $K(0)$ is 
below the SP prediction $1/2$, in agreement with our former numerical 
results on the number variance. 

A better way to compare the form factor to the SP statistics is 
introduced in Ref.~\cite{BGS01}. Fit $K(\tau;n)$ to the function 
\begin{equation}\label{eq:Kfit}
K_{\text{fit}}(\tau) = \frac{c^2-2c+4\pi^2\tau^2}{c^2+4\pi^2\tau^2} \ .
\end{equation}
If $c=4$ then function~(\ref{eq:Kfit}) is the SP form factor. We use
the quantity $K_{\text{fit}}(0)-1/2$ to measure the difference to the SP
statistics. Keep in mind that $K_{\text{fit}}(0)$, in general, differs from
$K(0;n)$ since it depends also on $K(\tau;n)$ with $\tau > 0$. 
Figure~\ref{fig:Kav} shows the result obtained by fitting \equ~(\ref{eq:Kfit})
to the smoothed data over the range $0\leq\tau\leq 3$. We get
$K_{\text{fit}}(0) \approx 0.552$. 
For the lower-energy windows with $k=600$, $k=1000$, and $k = 1300$ we find 
$0.548$, $0.576$, and $0.56$, respectively. 

Again, as for the other spectral quantities described in the previous
subsections, we find that the spectral statistics is roughly described by the
SP statistics. However, there are small but significant deviations which show
no clear trend to zero as the energies are increased.

\section{Conclusion}
\label{sec:conclusion}
We have shown how the scaling method can be applied to convex billiards with
corners. To gain insight into the strong diffractive phenomena that appear in
these systems we have studied the truncated triangle.
As a result of our investigations of the diffracted field we could 
identify its main components. These were directly introduced in the 
function basis in order to obtain the spectral data of this system. 
Evanescent waves conveniently selected 
and associated with the centers of diffraction (in our case the 
two straight segments junction at $3\pi/4$) have been 
successfully identified as a very efficient way to deal 
with these kinds of phenomena. This allowed us to obtain a great 
number of highly excited eigenvalues. 

We have studied the statistical properties of high-lying energy levels in this 
pseudointegrable billiard. We have found that the nearest-neighbor spacing 
distributions, next-to-nearest spacing distributions, number variance, and the
spectral form factor are roughly described by the semi-Poisson (SP)
statistics. Whether the SP statistics is asymptotically the exact statistics
cannot be decided.

\begin{acknowledgments}
We would like to thank Eduardo Vergini for his helpful suggestions 
during the writing of the manuscript. Also for the ideas he gave us 
regarding the evanescent wave approach to diffraction at corners.
Moreover, we would like to thank H.~Schomerus and T.~Gorin for discussions.
\end{acknowledgments}

\begin{appendix}
\section*{Appendix: The scaling method}
\label{sec:appendixA}

In this appendix we are going to briefly explain the scaling method~\cite{VerginiSaraceno95, PhD1,PhD2}. 
The main point that one should have in mind is that the boundary norm 
can be written as a function of energy because scaling is used. If 
$\phi(r)$ satisfies the Helmholtz equation with eigenvalue $k_0^2$ 
and we associate with it the scaling function $\phi(k,{\bf r}) = 
\phi(k {\bf r}/ k_0)$ then, these functions satisfy the same equation 
with eigenvalue $k^2$. If we have a billiard defined by a star-shaped 
domain and a $k_{\mu}$ exists such that
$\phi_{\mu}(k_{\mu}, {\bf r})=0$ at the boundary ${\cal C}$, then this is a
scaling eigenfunction or, equivalently, an eigenfunction of the billiard with
Dirichlet boundary conditions.  

The boundary norm defined by the expression $f(k)=\oint_{\cal C} \phi^2
(k, {\bf r}) dl /r_n$ can be expanded up to third order around $k_{\mu} $, 
independently on the exact shape of $\phi$.
We remind that $r_n={\bf r} \cdot {\bf n}$ (where ${\bf n}$ is the unit outgoing
normal 
vector to ${\cal C}$) is always positive for star-shaped domains. 
Taking into account this result we can evaluate the norm and its derivative 
in $k$ at value $k_0=k_{\mu}+ \delta_{\mu}$, obtaining
\begin{equation}
f(k_0)-\frac {\delta_{\mu}}{2} \frac{df}{dk}(k_0)+ {\cal O}(\delta_{\mu}^4)=0
\ .
\end{equation}

This useful expansion, dropping terms of order $\delta_{\mu}^4$, turns out to 
be our quantization condition. Then, all the scaling eigenfunctions with 
eigenvalues close to $k_0$ can be found by solving a 
generalized eigenvalue problem: 
\begin{equation}
\left[ \frac{dF}{dk} (k_0) - \lambda_{\mu} F(k_0) \right] \xi^{\mu} = 0 \ .
\end{equation} 
  
In this equation and for numerical calculations, the quadratic form 
$F$ associated with $f$ can be evaluated in a basis of scaling 
functions ${\psi_i(k, {\bf r});i=1,\ldots,N}$ (like plane waves for 
instance) by means of
\begin{equation}
F_{ij}(k_0) = \oint_{\cal C} \psi_i(k_0,{\bf r})\psi_j(k_0,{\bf r})dl/r_n \ .
\end{equation} 
The eigenfunctions are $\phi_{\mu}(k, {\bf r})= \sum_{i=1}^N \xi^{\mu}_i 
\psi_i(k, {\bf r})$, and the eigenvalues can be found as 
$k_{\mu}=k_0 - 2/ \lambda_{\mu}$. 
\end{appendix}

\bibliographystyle{prsty}
\bibliography{}

\end{document}